\documentclass[
 aip,
jcp,% short, numerical bibliography, 
 amsmath,amssymb,
 reprint,%
]{revtex4-1}

\usepackage[dvipdfmx]{graphicx}
\usepackage{dcolumn}% Align table columns on decimal point
\usepackage{bm}
\usepackage{color}
\newcommand\eq[1]{Eq.~(\ref{#1})}
\newcommand\Fig[1]{Fig.~\ref{#1}}

\newcommand\Sect[1]{Section~\ref{#1}}

\begin{document}

\preprint{}

\title[Hard-sphere melting and crystallization with event-chain Monte Carlo]{Hard-sphere melting and crystallization with event-chain Monte Carlo}

\author{Masaharu Isobe}
 \email{isobe@nitech.ac.jp}
\affiliation{Graduate School of Engineering, Nagoya Institute of Technology,
Nagoya, 466-8555, Japan}

\author{Werner Krauth}
 \email{werner.krauth@ens.fr}
\affiliation{Laboratoire de Physique Statistique, Ecole Normale
Sup\'erieure / PSL Research University, UPMC, CNRS, 24 Rue Lhomond, 75231 Paris
Cedex 05, France}


\begin{abstract}

We simulate crystallization and melting with local Monte Carlo (LMC),
event-chain Monte Carlo (ECMC), and with event-driven molecular dynamics (EDMD)
in systems with up to one million three-dimensional hard spheres. We illustrate
that our implementations of the three algorithms rigorously coincide in their
equilibrium properties. We then study nucleation in the NVE ensemble from the
fcc crystal into the homogeneous liquid phase and from the liquid into the
homogeneous crystal. ECMC and EDMD both approach equilibrium orders of
magnitude faster than LMC. ECMC is also notably faster than EDMD,
especially for the
equilibration into a crystal from a disordered initial condition at high
density. ECMC can be trivially implemented for hard-sphere and
for soft-sphere potentials, and we suggest possible applications of this
algorithm for studying jamming and the physics of glasses, as well
as disordered systems.
\end{abstract}

%\pacs{64.70.kj, 05.10.-a, 05.20.Jj}% PACS, the Physics and Astronomy
%Use showkeys class option if keyword
\maketitle

\section{Introduction}
Crystallization and melting have long been central subjects in statistical
physics.  These processes connect microscopic nucleation with the
macroscopic phenomena of domain growth and of phase transitions. A number of 
numerical
methods and simulation techniques have been brought to bear on these subjects,
following the pioneering computer simulations of hard-sphere systems
by both Monte Carlo~\cite{metropolis_1953,wood_1957,krauth_2006} and 
by molecular dynamics~\cite{alder_1957,alder_1959,alder_1962,alder_1968,alder_2009}.

The hard-sphere system is trivial to describe. Nevertheless,
equilibration in this simplest of all particle systems
is a slow process, because of the large activation free energy for
crystallization.
Timescales are also especially large in the fluid-solid coexistence regime,
because of the surface tension between coexisting phases. Specialized algorithms
for equilibration have been developed to overcome these problems, and the
melting and crystallization time scales provide useful benchmarks for their
comparison. 

A rejection-free hard-sphere ``event-chain'' Monte Carlo algorithm
(ECMC)~\cite{bernard_2009} has recently allowed to speed up equilibration for
two-dimensional hard disks by roughly two orders of magnitude compared to
the
event-driven molecular dynamics~\cite{alder_1959, isobe_1999} (EDMD) and to
local Monte Carlo~\cite{bernard_2012,engel_2013} (LMC). In ECMC,
a randomly sampled starting sphere moves along a straight line until the
latter
collides with another sphere, which then moves in the same direction
until it collides itself with
yet another sphere. This continues until the spheres'
total displacement equals a certain fixed length $L_c$. ECMC breaks detailed
balance
(moves are in the $+x$, $+y$, and $+z$ directions only) yet satisfies
global balance and ergodicity~\cite{michel_2014}. It rigorously samples the
equilibrium Boltzmann distribution. Considerable speedup was also demonstrated
for the extension of ECMC to continuous
potentials~\cite{michel_2014,kapfer_2014}.

In this paper, we assess the speed of ECMC, EDMD, and LMC not
by computing 
autocorrelation functions in equilibrium, but rather by the time scales 
associated with melting and crystallization in systems of many spheres
at high density. 
We expect our observations to extend to
subjects as dense packing, nucleation and jamming.
We focus on the melting from the metastable solid branch to 
the stable liquid (that is, slightly below the liquid--solid coexistence
interval) and also the nucleation processes from the metastable liquid branch 
towards the stable solid slightly above coexistence.
The paper is organized as follows: Our model system
and our methods are described in \Sect{s:model_algorithms_observables},
together with the observables on which we focus: The non-dimensional
pressure and the local and global orientational order parameters.
Results are summarized in \Sect{s:results}: We reproduce the phase diagram 
by ECMC and quantify the efficiencies of our three methods. 
We discuss the relative efficiency of ECMC and EDMD
for the crystallization process. Concluding remarks are
described in \Sect{s:conclusion}.
\begin{figure*}[ht]
\begin{center}
\begin{minipage}{0.42\hsize}
\includegraphics[scale=0.42]{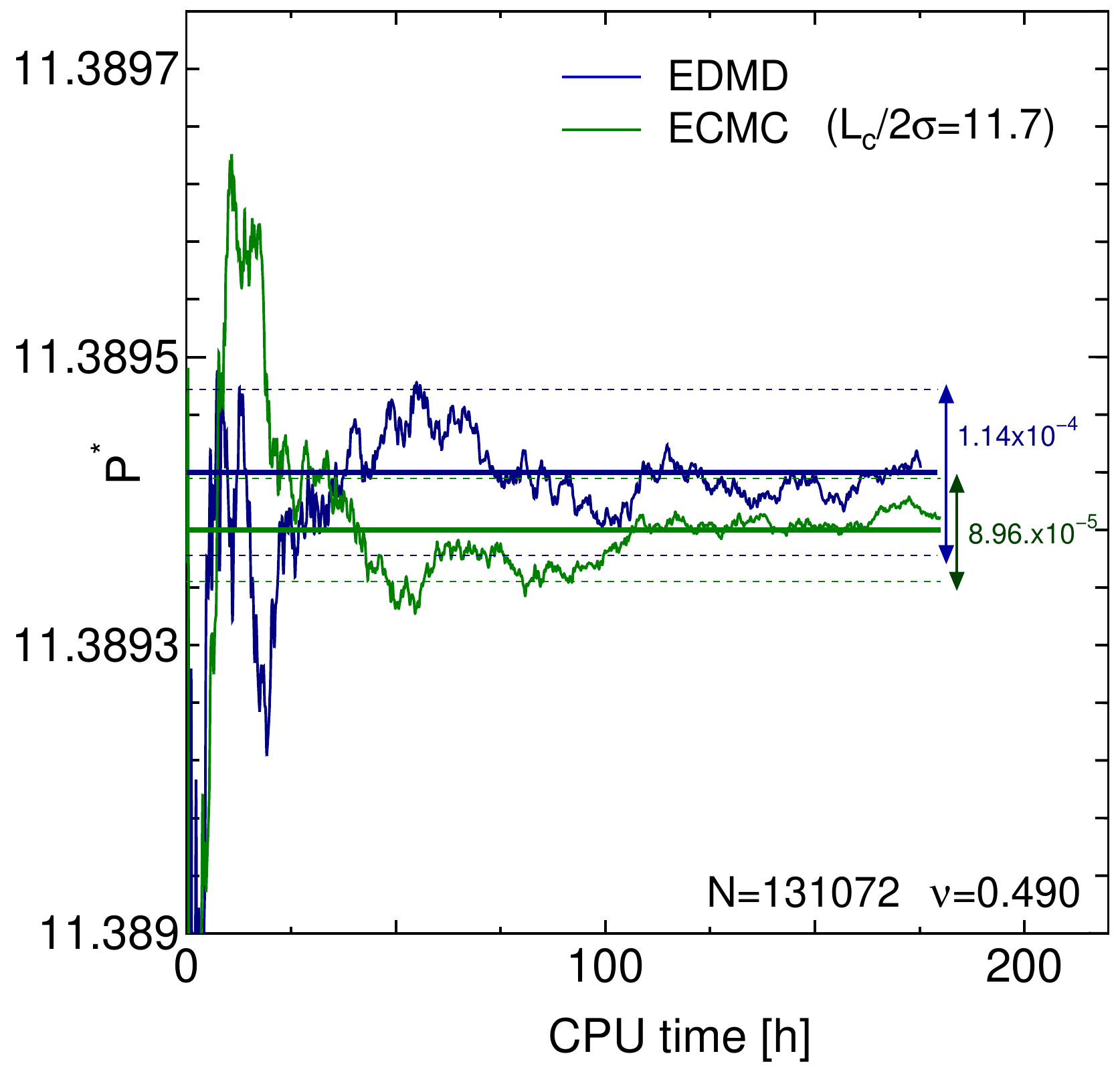}
\end{minipage}
\hspace{5mm}
\begin{minipage}{0.42\hsize}
\includegraphics[scale=0.42]{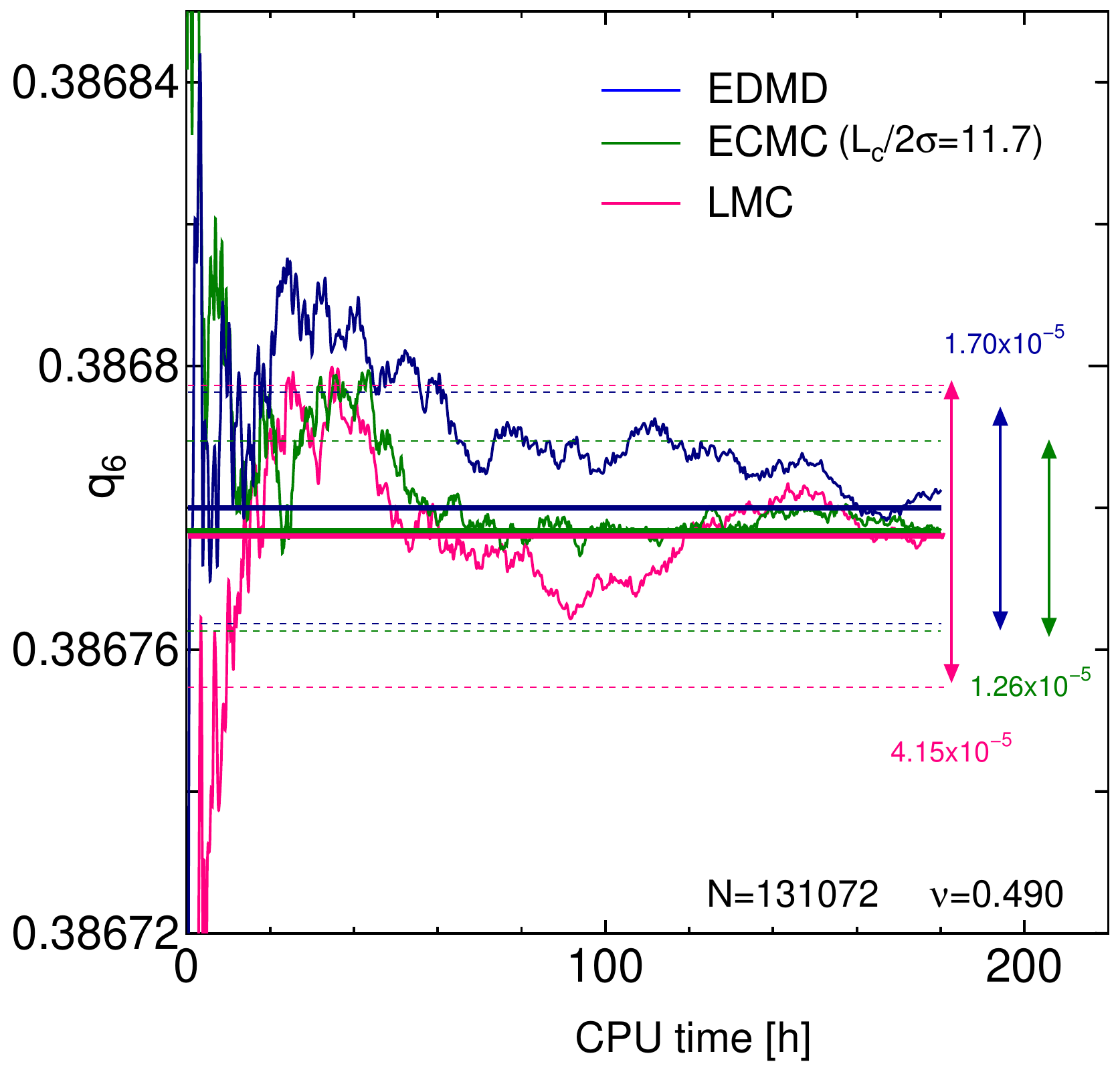}
\end{minipage}
\caption{
Pressure $P^*$ (left) and local order parameter $q_6$ (right) in the
dense liquid at $\nu = 0.490$, obtained from ECMC and EDMD for $N=131,072$ hard
spheres. In ECMC, the pressure is computed using the excess displacement method
of \eq{e:pressure_event}. In the right panel, the local order parameter of LMC
is also shown.}
\label{f:pressure_ECMC_EDMD}
\end{center}
\end{figure*} 

\section{Model, algorithms, and observables}
\label{s:model_algorithms_observables}

We consider $N$ monodisperse hard spheres of radius $\sigma$ in a cubic box of
sides $L$ with periodic boundary conditions (PBC). 
The density (packing fraction) $\nu$ is given by $\nu= 4/3 
N \pi \sigma^3/L^3$.
We concentrate on melting and crystallization from the unstable to the stable
phase, i.e., from the unstable crystalline branch to the liquid and
from the unstable liquid branch to the crystal.
For our melting runs, at density $\nu = 0.490$ below the coexistence
interval of liquid and solid phases, we
prepare the initial configurations as perfect fcc crystals, corresponding to
the stable phase at high densities (the free energy difference between fcc and
hcp crystal have been discussed actively since the works of
Ref.~\cite{woodcock_1997,bolhuis_1997}). The fcc initial conditions are
compatible with the cubic simulation box.

For the crystallization runs
at density $\nu =  0.548$ above the coexistence interval,
we start from a fluid initial configuration at a liquid-phase
density $\nu= 0.490$. In order to reach the higher target density, we repeatedly
increase $\sigma$ slightly and remove all created overlaps by sliding
overlapping pairs of spheres for a half length of overlap along their 
common symmetry axis. This is done until all pair overlaps
have disappeared.

In EDMD, hard spheres evolve in continuous physical time through collisions, 
and the dynamics solves Newton's classical equations of motion. 
We use an efficient sequential implementation~\cite{isobe_1999}. 
LMC and ECMC are implemented very simply. 
For the former, the optimal displacement of spheres is determined by short LMC
runs from the initial conditions so that the acceptance ratio is 1/2.
For the latter, a single parameter, the chain length $L_c$, must be optimized
for each density. 
A single event can be implemented very quickly in ECMC, as the motion is 
always in +x or +y, which decreases the CPU time per event.
We consider systems with $N=2,048$, $131,072$ and $1,048,576$
spheres. Our calculations are mainly done on the Intel Xeon CPU E5-2680 2.80GHz,
where we reach $\sim 3.15 \times10^9$ events/h of CPU time for ECMC and $\sim
4.62 \times10^8$ collisions/h for EDMD. For the LMC algorithm, $\sim 6.5
\times 10^9$ trials/h are reached. 
All our comparisons of algorithms are in
terms of CPU time.
To be as fair as possible, the three algorithms were implemented 
following unified design principles. Furthermore, we used 
the same computer, the same language (Intel FORTRAN), and the same optimal 
option of compiler. An event-count would have produced similar results.

We track the time-evolution of the ordering/disordering of the system from the
pressure and the local and global orientational order parameters. In EDMD, the
nondimensional virial pressure is computed from the collision rate via the
virial theorem:
\begin{equation}
P^* = \beta P (2\sigma)^3 = \frac{6 \nu}{\pi} \left[ 1-\frac{\beta
m}{3T}\frac{1}{N} \sum_{\rm collisions} b_{ij} \right],
\label{e:pressure_md}
\end{equation} 
\noindent
where $T$ is the total simulation time, and $\beta=1/m\left<v_x^2\right>$ is 
the inverse kinetic temperature (mass $m$ and mean-square $x$-component of
velocity of spheres). 
The collision force $b_{ij} =  {\bf r}_{ij}\cdot {\bf v}_{ij}$ is defined
between the relative positions and the relative velocities of the collision
partners~\cite{engel_2013, erpenbeck_1977}.
In ECMC, the pressure $P^*$ can be evaluated from the mean excess chain
displacement~\cite{michel_2014}: 
\begin{equation}
P^* = \frac{6 \nu}{\pi} \left< 
\frac{x_\text{final}-x_\text{initial}}{L_c} \right>_\text{chains}, 
\label{e:pressure_event}
\end{equation}
where $x_\text{final}$ and $x_\text{initial}$ are final and initial positions of
each chain, respectively, taking into account the PBC. 
This convenient formula replaces the tedious
extrapolation of the pair correlation function at contact $g_2(r = 2 \sigma)$
that was used previously and that must still be used for LMC.
The comparison of the evolving pressure of
EDMD (using Eq.~(\ref{e:pressure_md}))
and ECMC (using Eq.~(\ref{e:pressure_event}) in the liquid state at $\nu=0.490$
is shown in the left of \Fig{f:pressure_ECMC_EDMD}. The pressures fluctuate
around $P^*=11.3893$, and agree within very tight error bars.

Besides the pressure, we  quantify the speed of melting and of crystallization
via the time-dependent local $q_6$ and global $Q_6$ order
parameters~\cite{steinhardt_1983}:
\begin{eqnarray}
q_6 & = & \frac{1}{N} 
     \sum_{i=1}^N \sqrt{\frac{4\pi}{13}
     \sum_{m=-6}^{m=6} \left| \frac{1}{n(i)}\sum_{j=1}^{n(i)} Y_{6,m} ({\bf
r}_{ij})\right|^2}, \\
Q_6 & = & \sqrt{\frac{4\pi}{13}
     \sum_{m=-6}^{m=6} \left| \frac{1}{N n(i)} \sum_{i=1}^N 
     \sum_{j=1}^{n(i)} Y_{6,m} ({\bf r}_{ij})\right|^2},
\end{eqnarray}
where $n(i)$ is the number of nearest neighbors for each sphere
$i$ and $Y_{6,m}({\bf r}_{ij})$ are the spherical harmonics with icosahedral
symmetry for the distance vector ${\bf r}_{ij}$ between spheres $i$ and $j$.
We detect nearest neighbors by the SANN algorithm of Meel et
al.\cite{meel_2012} rather than by the Voronoi construction. On the right of
\Fig{f:pressure_ECMC_EDMD}, we again demonstrate that the equilibrium values for
$q_6 (\sim 0.38678)$ agree within tight error bars for LMC, ECMC and EDMD.
Similar agreement was reached for the global order parameter $Q_6$.
We note that perfect fcc configurations have an orientational order $Q_6 = q_6 
\sim 0.575$, while for liquid configurations (including our disordered initial 
configurations) $Q_6$ approaches zero. In the dense liquid, the local order
$q_6$ is non-zero because of the build-up of transient local crystal
structures~\cite{isobe_2012}.

\section{Results}
\label{s:results}

\subsection{Hard-sphere phase diagram}
\label{s:phase_diagram}
\begin{figure}[ht]
\begin{center}
\includegraphics[scale=0.5]{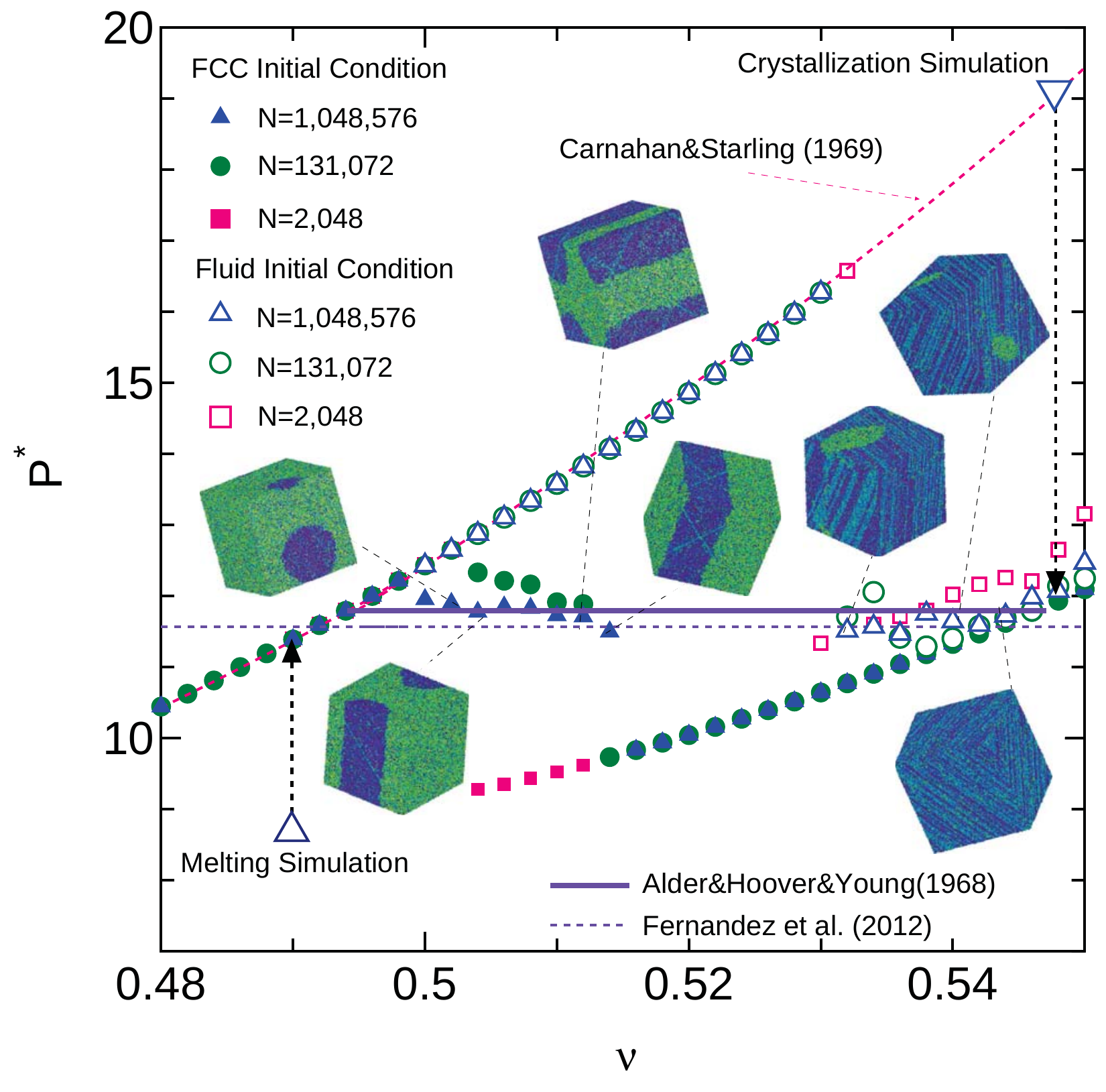}
\end{center}
\caption{
Hard-sphere equation of state obtained by ECMC  after $\sim {\cal  O}(10^{12})$
collisions. The snapshots in the coexistence interval at $N=1,048,576$
represent the local orientational order $q_6 (i)$ for each sphere $i$
(liquid-like local order is represented in green - solid-like local order 
in blue). 
The equilibrium coexistence pressure for the infinite system is also shown.
The densities $\nu = 0.490$ and $\nu = 0.548$, on which we concentrate in
Sections~\ref{s:melting} and \ref{s:crystallization}, are indicated. 
}
\label{f:equation_of_state} 
\end{figure}

The fluid-solid coexistence in the $NVE$ ensemble (which, for hard spheres, 
corresponds to the common $NVT$ ensemble) 
for densities $\nu$ in the interval $0.494 < \nu < 0.545$ has 
been investigated for more than 50 years~\cite{alder_1957,wood_1957,alder_1968}. \begin{figure}[hb!]
\includegraphics[scale=0.42]{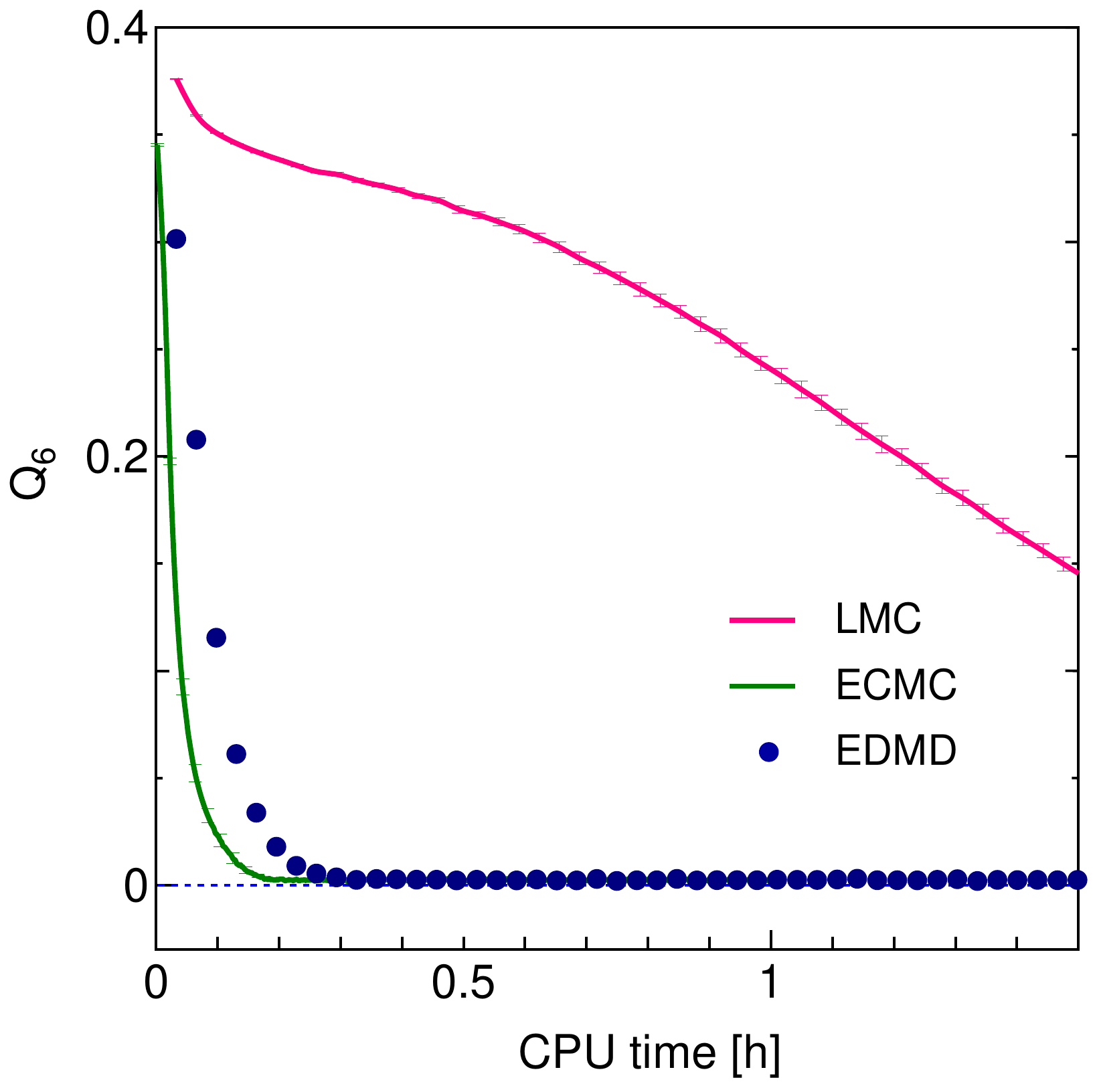}
\caption{Melting at $\nu=0.490$ ($N=131,072$) from an fcc initial configuration
into the stable liquid, tracked by the time evolution of the global $Q_6$ order
parameter in LMC, EDMD, and ECMC with optimal chain length 
($L_c/2 \sigma = 5.87$). 
Data averaged over 5 samples. Note that ECMC and EDMD are orders 
of magnitude faster than LMC.}
\label{f:Q6_melting}
\end{figure}
Recently, various theoretical equations of states were compared with the results
of  a large-scale EDMD simulation on this system,~\cite{bannerman_2010}
with $N \sim 10^6$.
The metastable fluid branch in the  fluid-solid coexistence window was
found stable against freezing
on EDMD time scales up to $\sim {\cal O}(10^9)$ collisions.
To speed up the simulation, the replica exchange MC method was adapted to the
hard-sphere case. To keep the acceptance rate at reasonable values,
many replicas at finely spaced densities had to be used, and 
this approach proved restricted to quite small system sizes ($N=32$
and $N=108$).~\cite{odriozola_2009} Fernandez et
al.~\cite{fernandez_2012} explored the coexistence of hard-sphere systems
in equilibrium by tethered MC for relatively small system sizes $\sim {\cal
O}(10^3)$. In this method, the approach to equilibrium is accelerated
by a biased field of two order parameters. The equilibrium pressure
$P^*=11.5727(10)$ is obtained through extrapolation towards the infinite-size
limit. We note that in three-dimensional hard spheres, the direct simulation
remains difficult in the coexistence region, even for ECMC, whereas for the
analogous two-dimensional hard disks, the equilibration of the
coexisting hexatic and liquid phases by ECMC proved possible at all densities,
for up to one million disks.\cite{bernard_2011}
\begin{figure*}[ht!]
\begin{center}
\begin{minipage}{0.42\hsize}
\includegraphics[scale=0.42]{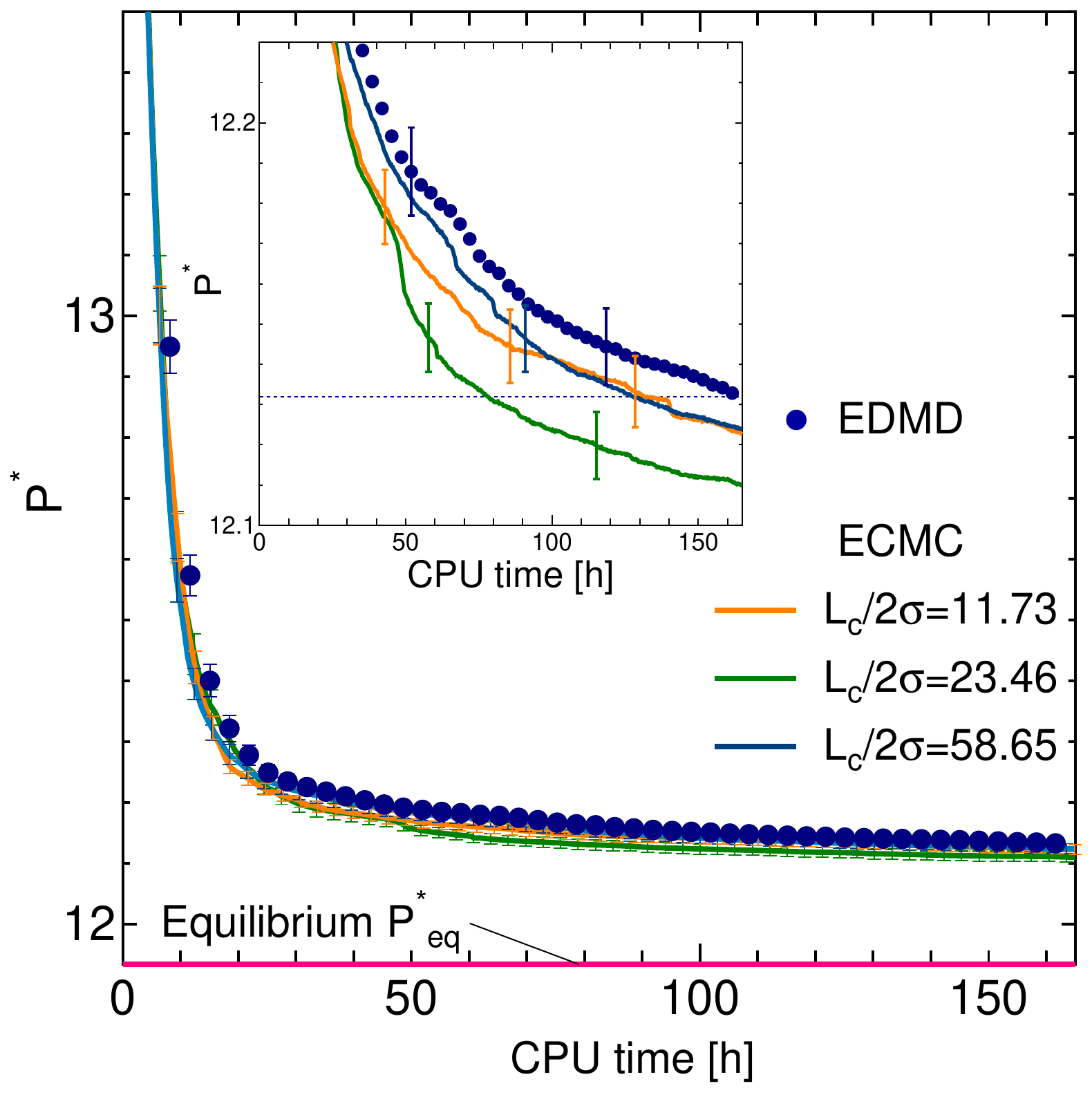}
\end{minipage}
\hspace{3mm}
\begin{minipage}{0.42\hsize}
\includegraphics[scale=0.42]{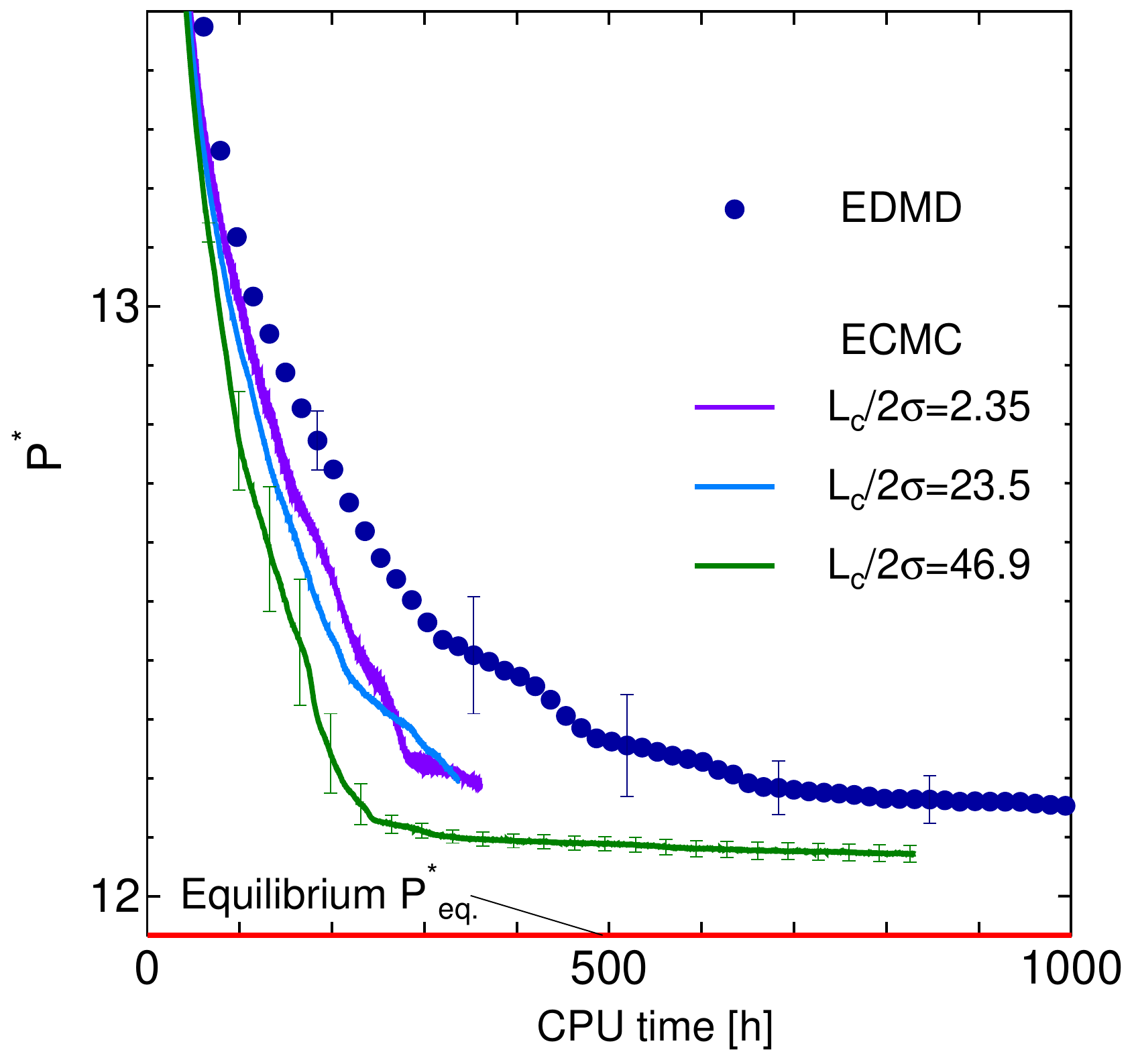}
\end{minipage} 
\end{center}
\caption{ Crystallization at $\nu=0.548$ from a random initial configuration,
tracked by the evolution of the pressure $P^*$  in EDMD
and ECMC with different chain lengths $L_c$.
Data averaged over $100$ samples for
$N=131,072$ (left) and $5$ samples for $N=1,048,576$ (right). 
The inset in the left panel illustrates the influence 
of the parameter $L_c$ on the performance of ECMC.}
\label{f:press_crystal} 
\end{figure*}

\Fig{f:equation_of_state} shows the hard-sphere equation of state from
ECMC (final pressures after long runs with ($2 \sim 3 \times 10^{12}$)
collisions). The pressure is averaged over $10^{10}$ collisions at the
end of the simulation.
The stable and unstable fluid pressures well agree with 
Hoover-Ree~\cite{hoover_1968} and Carnahan-Starling
extrapolation.~\cite{carnahan_1969} Inside the coexistence interval,  
the final pressure depends on the initial configuration, as the metastable 
fcc solid or fluid initial configurations are preserved on the time-scale of
the simulation. In the $NVE$ ensemble, the presence of interfaces of different
topologies makes that the equation of state is non-monotonous, and the liquid-solid
coexistence pressure curve is not flat in a finite system. As one increases the
density from the liquid phase, the spherical or cylindrical droplets that can be
seen in \Fig{f:equation_of_state} generate an excess Laplace
pressure. This is analogous to what was found in
two-dimensional hard disks~\cite{bernard_2011,mayer_1965}, which show 
droplets and stripes or in fluid-gas mixtures of the three-dimensional 
Lennard-Jones system~\cite{Binder_AmJP}, where spherical and cylindrical 
droplets as well as two-dimensional stripes are found.

Specifically, for $\nu < 0.498$, simulations from arbitrary
initial conditions converge to the same pressure since the system is
completely liquid and nucleation barriers are low.  In the
region of $\nu = 0.500 \sim 0.514$,
simulations from fcc initial configurations successfully create
interfaces with different topologies, 
whereas simulations from fluid initial conditions remain on the fluid branch.
The pressure at $\nu=0.498 \sim 0.514$ decreases as $P^*= 12.2 \sim 11.5$,
and agrees with the expected coexistence pressure.
\cite{fernandez_2012}
The phase coexistence at equilibrium can be seen clearly by the spatial
distributions of the local $q_6$ order parameter, where the fcc crystal
reduces to a droplet ($\nu=0.500 \sim 0.502$) when started from an
fcc
crystal. For larger densities ($\nu=0.504 \sim
0.512$) the remaining fcc phase has the form 
of a cylinder that reconnects through the PBC, surrounded by the liquid dominant
phase created through melting (see the insets of \Fig{f:equation_of_state}).  
In the density interval $\nu = 0.514 \sim 0.530$, the fcc crystal and the liquid
remain metastable on the available scales of simulation time.  For $\nu = 0.532
\sim 0.543$, simulations from fluid initial conditions nucleate fcc droplets.
At $\nu > 0.543$, simulations from fluid initial condition drop down near 
the solid branch, however, relaxation is in progress at the value around 
slightly higher pressure than the solid branch, except for the case of 
$N=2,048$. 

\subsection{Melting from an fcc initial configuration at a fluid density}
\label{s:melting}

We now study the speed of melting into the equilibrium liquid phase at $\nu =
0.490$ from an fcc crystalline initial configuration for $N = 131,072$.  From
the global order parameter $Q_6$, as shown in \Fig{f:Q6_melting}, the initial
fcc configuration ($Q_6 \sim 0.575$) rapidly becomes unstable for the three
algorithms and approaches the liquid branch, where the global orientational
order approaches zero. Melting is much faster for ECMC and EDMD than for LMC.
ECMC is found to be somewhat faster than EDMD.
\begin{figure*}[ht!]
\begin{center}
\begin{minipage}{0.42\hsize}
\includegraphics[scale=0.42]{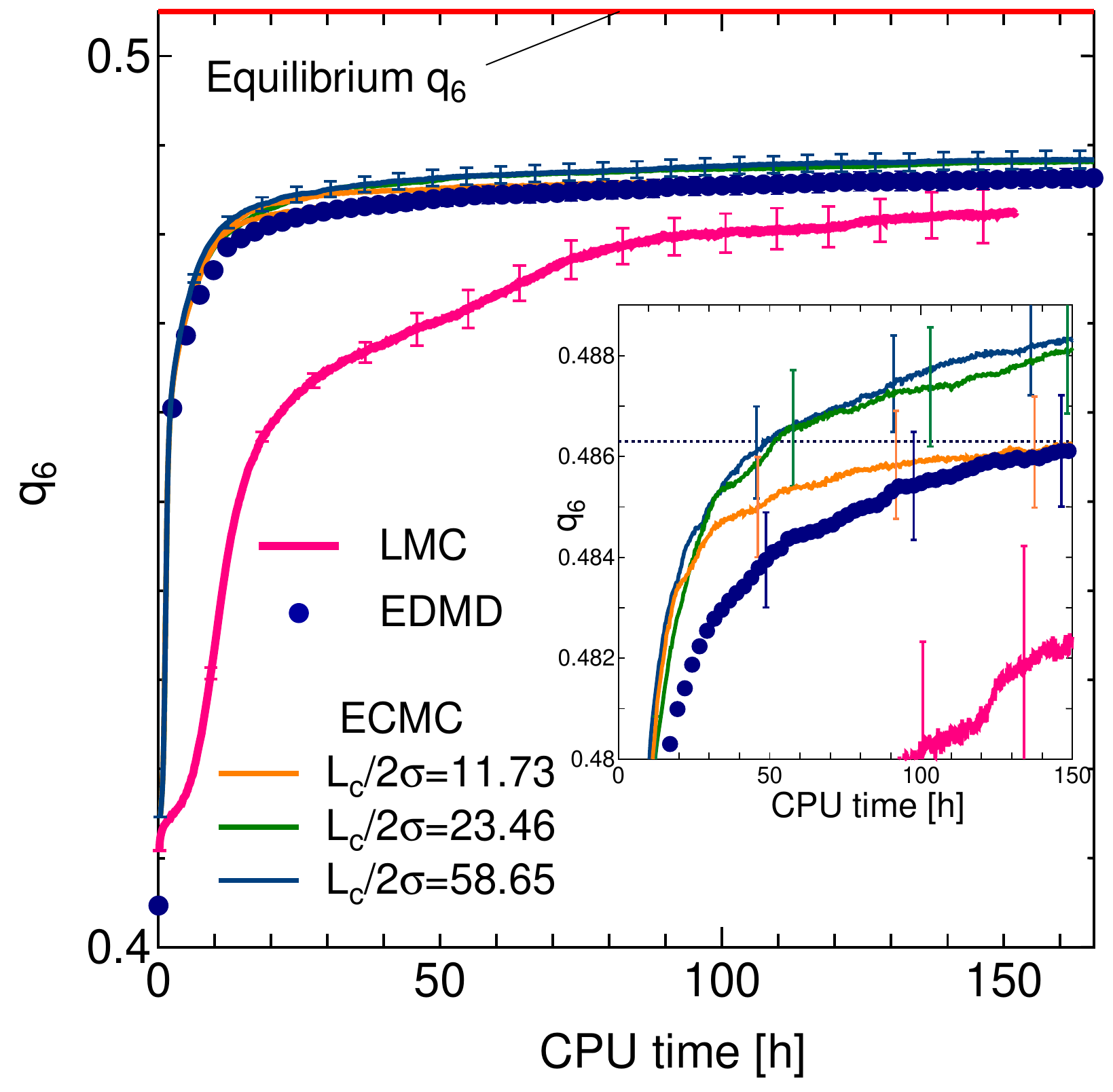}
\end{minipage}
\hspace{3mm}
\begin{minipage}{0.42\hsize}
\includegraphics[scale=0.42]{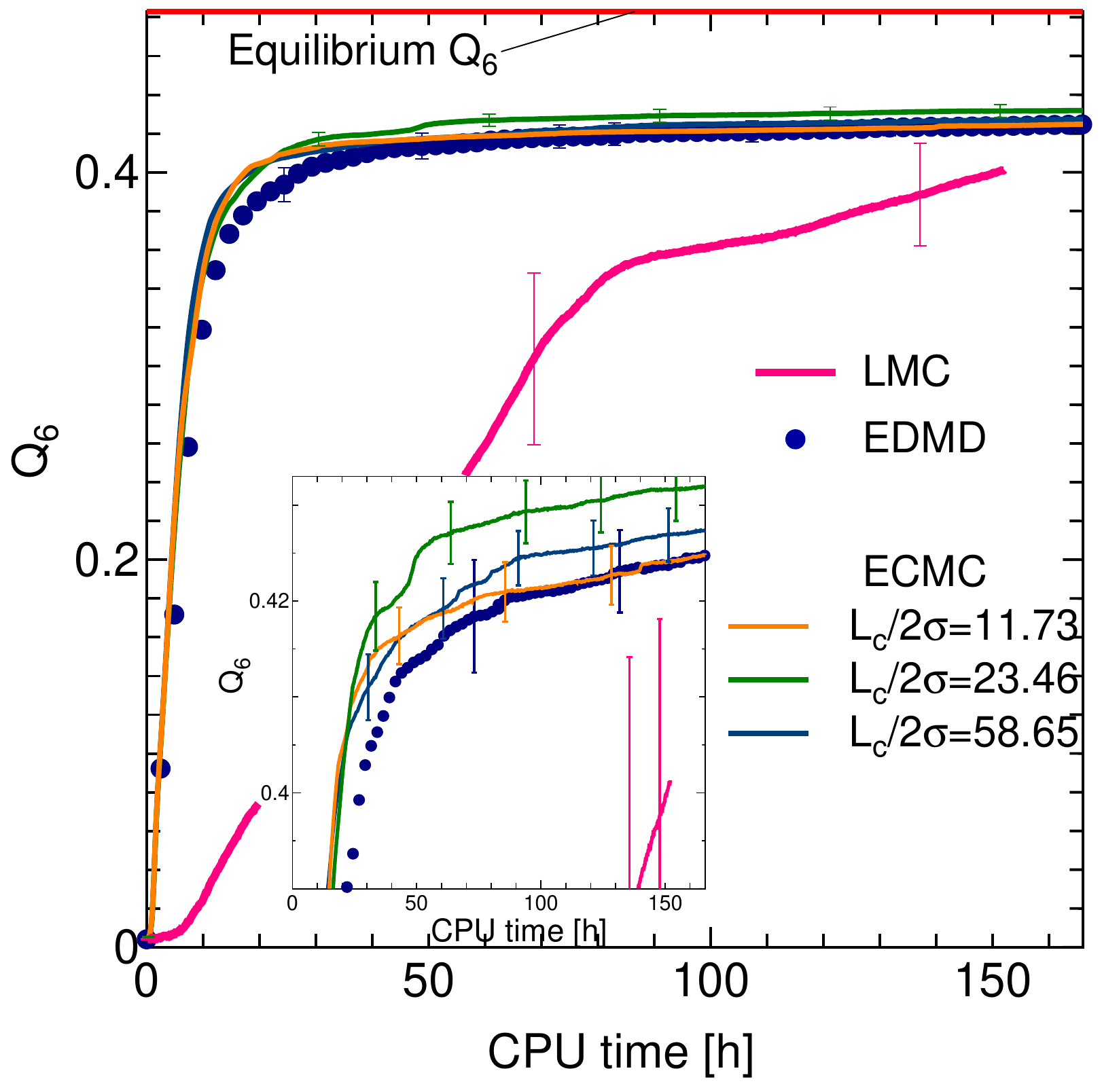}
\end{minipage}
\end{center}
\caption{
Crystallization at $\nu=0.548$ from a fluid initial configuration with $N = 
131,072$, tracked by the evolution of local $q_6$ and global $Q_6$ order
parameters by LMC, EDMD and ECMC with different chain lengths $L_c$. Data
averaged over $100$ samples.
}
\label{f:q6_crystal_small} 
\end{figure*}

\subsection{Crystallization from a fluid initial condition at a solid density}
\label{s:crystallization}

In \Fig{f:press_crystal}, we show the evolution toward crystallization of EDMD
and of ECMC with different  chain lengths $L_c$. Results are averaged over $100$
samples for $N=131,072$ (left of \Fig{f:press_crystal})
and over $5$ samples for $N=1,048,576$ (right of \Fig{f:press_crystal}).
Results for three trial runs in which we changed the 
chain length $L_c$ are also shown \Fig{f:press_crystal}. 
The efficiency of ECMC naturally depends on $L_c$.
For both methods, the pressure
remains somewhat above the configurational equilibrium pressure $P^* \sim
11.934$. The relative advantage of ECMC with optimal chain length is evident.

In \Fig{f:q6_crystal_small} and \Fig{f:q6_crystal_large}, we show the evolution
of the local $q_6$ and global $Q_6$ order parameters in crystallization runs 
of EDMD and ECMC with three chain lengths.  The number of
samples is again $100$ for $N=131,072$ and $5$ for $N=1,048,576$.
In the early stage of relaxation, $q_6$ and $Q_6$ are increasing functions of
CPU time.
In the perfect fcc configuration,
the local and global order parameters agree to $q_6 = Q_6 = 0.57452$.
Due to thermal fluctuations, actual numerical simulation at fcc equilibrium
estimate the order parameters to $(q_6, Q_6)=(0.505,0.483)$ in
$(N,\nu)=(1,048,576,0.548)$. 
Although the crystallization still proceeds, our final
averaged $q_6$ and $Q_6$ reach around $0.488$ and $0.432$, respectively.
The inconsistency between
the fcc crystal structure and the (finite) simulation box may well be
responsible for the reduction in order parameter.

Order parameters decay towards higher
order in time, ECMC with optimal chain length $L_c/(2\sigma)=23.46$ needs $48.6$
CPU hours to reach at $q_6=0.4863$ in $N=131,072$, however, EDMD needs $165$ CPU
hours.  Both smaller and larger $L_c$ results in the inefficiency of relaxation.
After $165$ hours, one quarter of ECMC simulations had almost reached 
the equilibrium state (i.e, $q_6 > 0.99 \times {q_6}_{eq.}$) whereas for
EDMD, only $10$ \% had reached such values.
The results for LMC are also shown in \Fig{f:q6_crystal_small}, and they show 
that it is much slower than ECMC and EDMD.  For example, our LMC
simulation on average reaches $q_6=48.25$ after ${\cal O}(10^7)$ LMC sweeps
corresponding to $152$ CPU hours.  At this $q_6=48.25$, ECMC and EDMD needs only
about $11$
and $28$ CPU hours, respectively.  The decay of
$Q_6$ has the same tendency as that of $q_6$, however, its values are lower
than that of $q_6$.  Since the whole system is slower to order
than
to establish local order, the global orientation also grows  more slowly
than the 
local orientation.  In the larger case $N=1,048,576$, ECMC with optimized chain
length and
EDMD need $416$ and $1,000$ CPU hours to reach $q_6 = 0.4897$,
respectively.  Note that if chain length is not optimized, the performance
of ECMC changes drastically and becomes comparable to or slower
than that of EDMD.  As for $q_6$ and $Q_6$, ECMC
with optimal length is faster than that of EDMD for a certain factor 
depending on the target point, which will be discussed in
\Sect{s:relative_speed}.

\subsection{Relative Speed of ECMC and EDMD}
\label{s:relative_speed}

To further quantify the equilibration speed of ECMC and EDMD,
rather than the observables as a function of time ${\cal O}(t)$,  we consider
the elapsed CPU time ${\cal T}({\cal O})$ from the beginning of the simulation
$t=0$ at which the observable ${\cal O}$ is reached.
This allows us to define the relative efficiency $R_s$:
\begin{equation}
R_s ({\cal O})= {\cal T}_{\rm EDMD}({\cal O})/{\cal T}_{\rm ECMC}({\cal O}), 
\label{e:relative_speed}
\end{equation}
and analogously for any pair of algorithms, where ${\cal O}$ is an observable,
in our case the pressure $P^*$, or the local and global order parameters $q_6$
and $Q_6$.
\begin{figure*}[ht!]
\begin{center}
\begin{minipage}{0.42\hsize}
\includegraphics[scale=0.42]{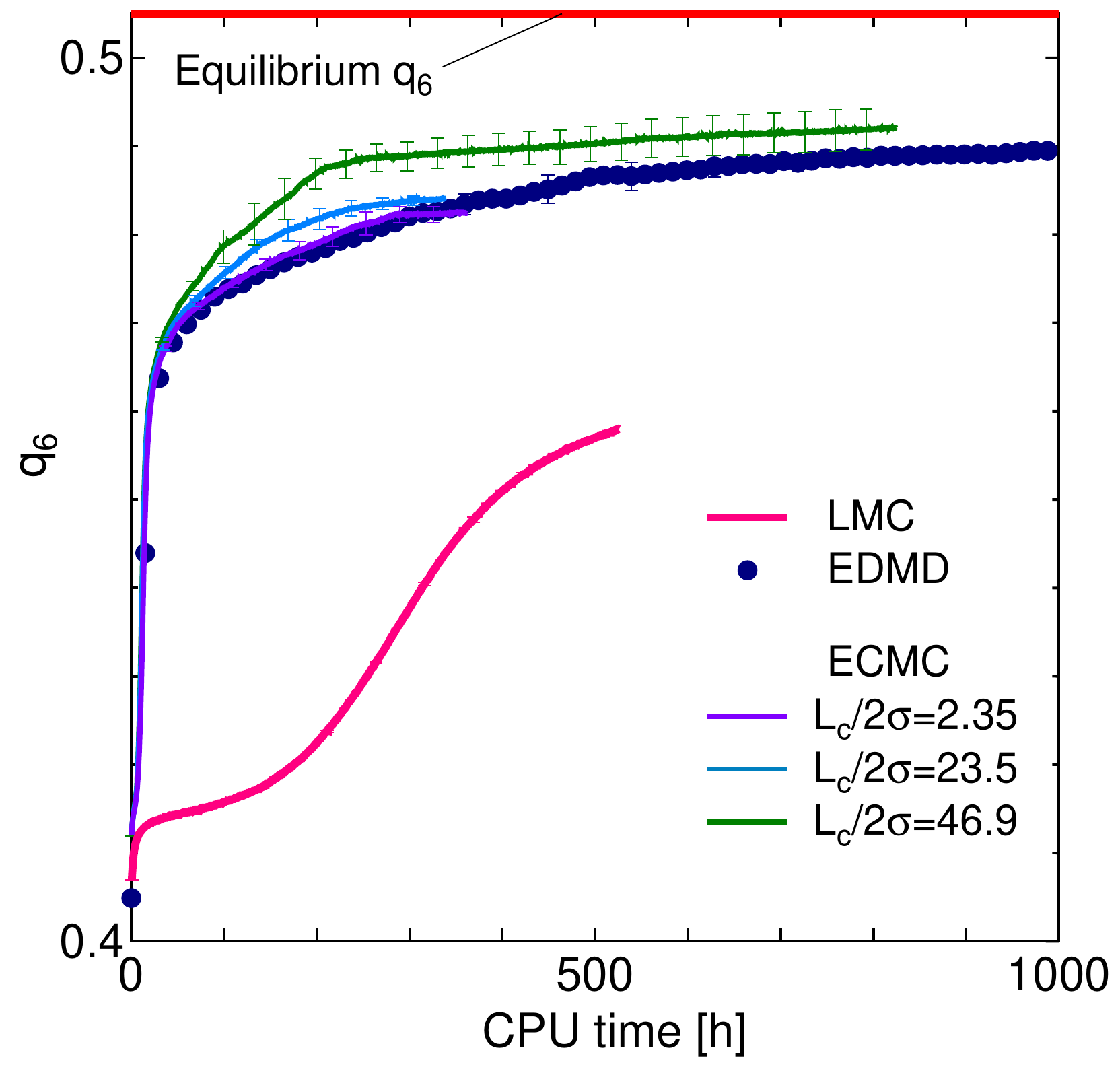}
\end{minipage}
\hspace{3mm}
\begin{minipage}{0.42\hsize}
\includegraphics[scale=0.42]{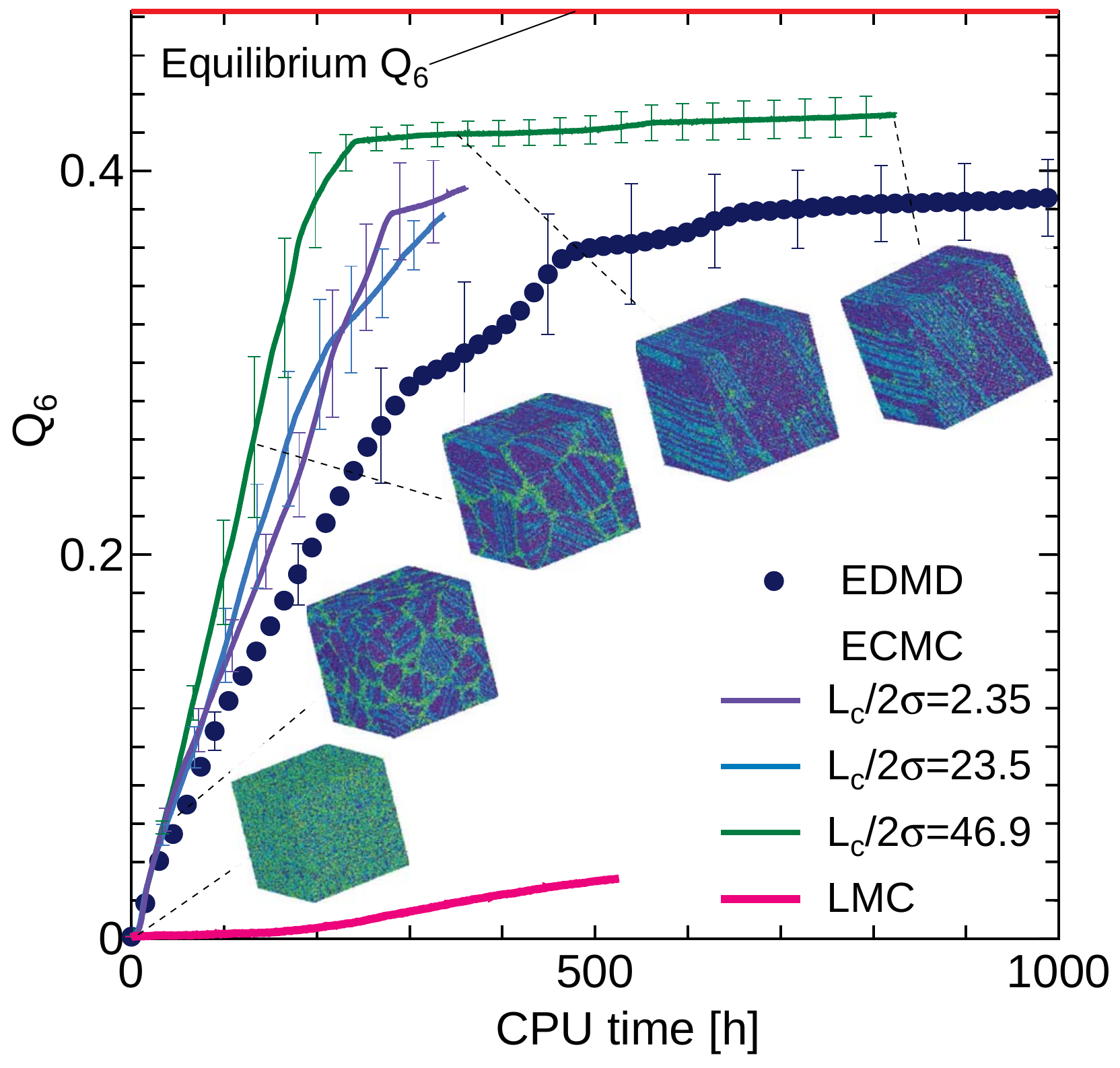}
\end{minipage}
\end{center}
\caption{
Crystallization at $\nu=0.548$ from a random initial configuration with
$N = 1,048,576$ tracked by
the evolution of local $q_6$ and global $Q_6$ order parameters in LMC, EDMD 
and ECMC with different chain lengths $L_c$. Data averaged over $5$ samples. 
Note that LMC is much slower than the other two methods. }
\label{f:q6_crystal_large} 
\end{figure*} 

\begin{figure*}[ht!]
\begin{center}
\begin{minipage}{0.42\hsize}
\includegraphics[scale=0.42]{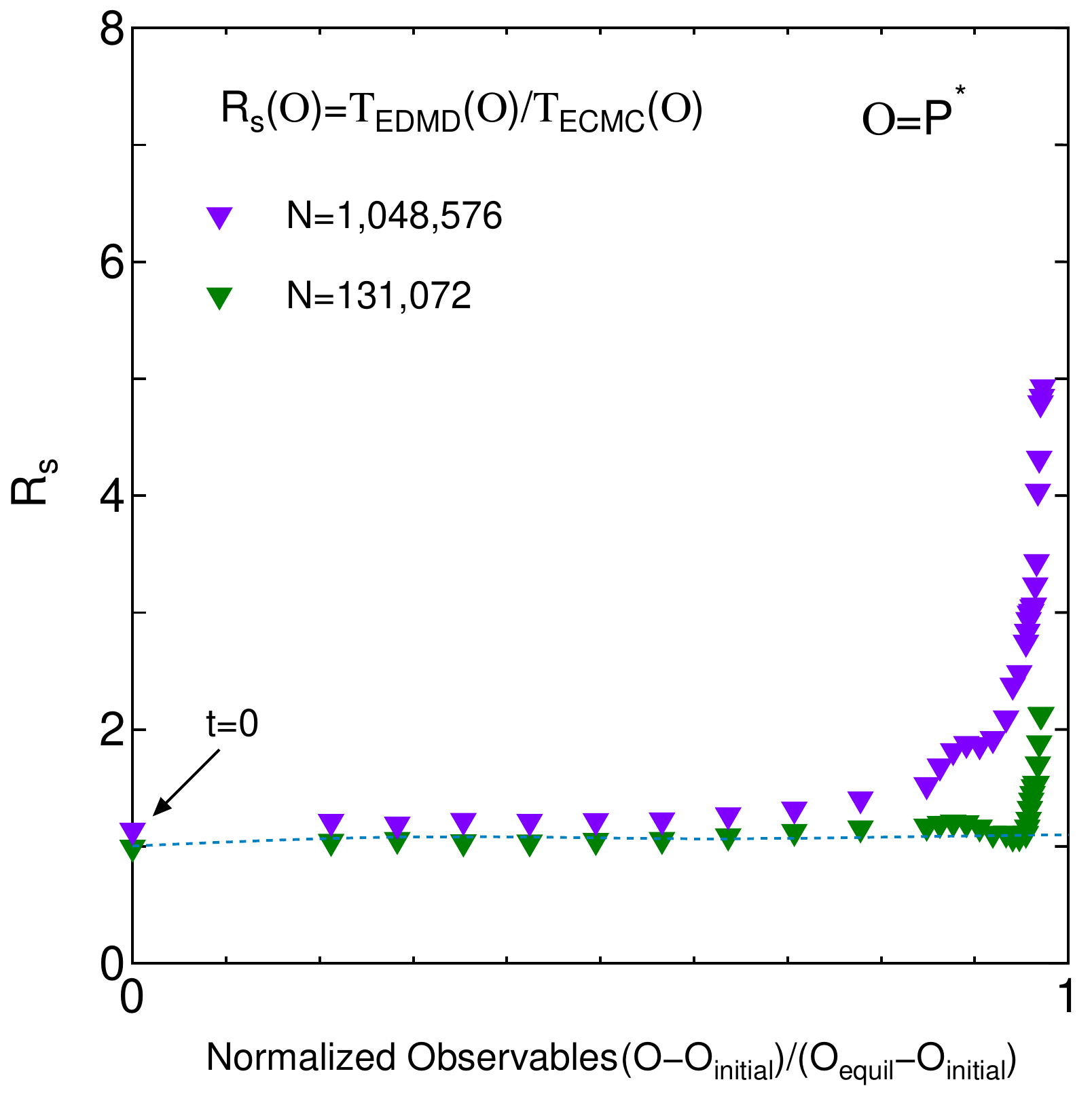}
\end{minipage}
\hspace{3mm}
\begin{minipage}{0.42\hsize}
\includegraphics[scale=0.42]{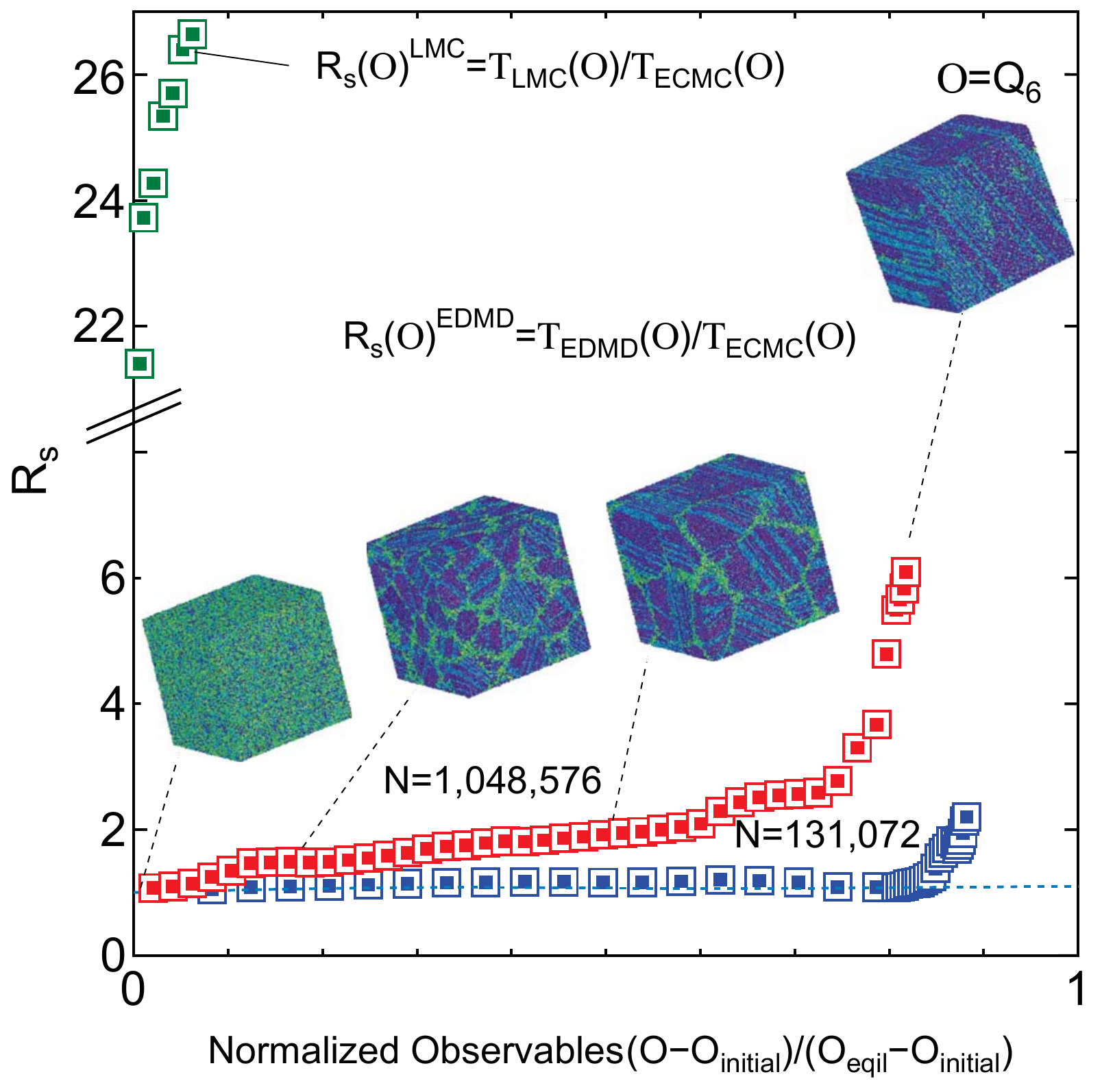}
\end{minipage}
\end{center}
\caption{ Relative speed $R_s$ (see \eq{e:relative_speed}) of LMC, ECMC and
EDMD as a function of normalized observables $\hat{\cal O}=({\cal O}-{\cal
O}_{\text{initial}})/({\cal O}_{\text{equil.}}-{\cal O}_{\text{initial}})$, 
(left) ${\cal O}=P^*$ and (right) ${\cal O}=Q_6$ for $N=1,048,576$.
As also shown in \Fig{f:q6_crystal_large}, both ECMC and EDMD are
orders of magnitude faster than LMC. ECMC shows considerable advantage over
EDMD in the later times of the evolution, when large-scale structures are 
built up.}
\label{f:relative_speed}
\end{figure*}

For the melting case (not shown) at $\nu=0.490$, $R_s$ for observables takes
around $3.2$ ($t=0$) to $1$ at the end of simulation. 
EDMD is slightly slower than that of ECMC. This relative speed remains rather
unchanged during the melting process.
The situation changes drastically for the crystallization process. 
\Fig{f:relative_speed} shows $R_s$
as a function of normalized observables as $\hat{\cal O}=({\cal O}-{\cal
O}_{\text{initial}})/({\cal O}_{\text{equil.}}-{\cal O}_{\text{initial}})$ at
$\nu=0.548$ in the crystallization process, at which each data can be estimated
by \Fig{f:press_crystal} - \Fig{f:q6_crystal_large}. 
${\cal O}_{\text{initial}}$ and ${\cal O}_{\text{equil.}}$ are the values 
at $t=0$ and at the equilibrium, respectively. Those are obtained by 
independent runs at the perfect fcc crystal which are $q_6=0.505$, 
$Q_6=0.483$, and $P^*=11.934$. 
In all cases, $R_s ({\cal O})$ is increasing function and growing drastically 
near the crystal (i.e., $\hat{\cal O} > 0.8$) as a {\em hockey stick} curve. 
The relative efficiency $R_s$ depends rather weakly on system size. 
We did not compute variations precisely, as the running times ${\cal T}$ 
were only averaged over 5 samples for $N = 1,048,576$. The different 
algorithmic complexity of our methods might marginally contribute to 
the size dependence of $R_s$: Our EDMD algorithm is implemented in 
$O(N \log N)$ per $N$ collisions and ECMC as $O(N)$ per $N$ collisions.
At $q_6=0.47$, $R_s(q_6)$ takes $1.33$ ($N=131,072$) and $1.49$ ($N=1,048,576$),
respectively.

\section{Conclusion}
\label{s:conclusion}

In this paper, we compared hard-sphere Monte Carlo and Molecular Dynamics 
algorithms, that coincide in their equilibrium properties. 
In large systems with up to one million
spheres, we recovered the known phase diagram and especially the coexistence
region.
We quantified the approach towards equilibrium, namely towards the
fcc crystal from the liquid-like initial configuration at packing $\nu = 0.548$
or the stable liquid from an fcc initial configuration at packing $\nu = 0.490$.
We clearly showed that the EDMD and ECMC are orders of magnitude
faster than the LMC algorithm for both the melting and the
crystallization.
ECMC needs optimization for chain length $L_c$, and we generally find
that the individual chains should wrap a few times around the simulation box.
The effect of the chain length is rather drastic, and $L_c$ must be
optimized carefully.
The optimal chain length for the crystallization process is estimated around 
$L_c/(2\sigma) \sim 25$ ($N=131,072$) and $50$ ($N=1,048,576$).
With a fixed $L_c/(2\sigma)$, the actual chain length 
$\left<x_{\text{final}}-x_{\text{initial}}\right>/(2\sigma)$ can be obtained by
trial and error before the production runs. It may also be estimated by the
ECMC pressure formula, \eq{e:pressure_event}), as
\begin{equation}
\frac{\left<x_{\text{final}}-x_{\text{initial}}\right>}{2\sigma}=\frac{P^* (L_c/(2\sigma))\pi}{6\nu}.
\end{equation}
which is evolving during simulation according to pressure relaxation.
In case of the optimal chain length $L_c/2\sigma = 23.46 \sim L_x/2$, 
the chain winds around $6$ times around the periodic box. While doing so, 
very few spheres get hit more than once.

We conclude that ECMC with well-chosen chain lengths is far superior 
to LMC, although it can be implemented just as easily~\cite{kapfer_2013, kampmann_2015}.
Even with respect to molecular dynamics, it performs very well.
The clearest advantage of ECMC over EDMD shows up in the crystallization, that
is, in the buildup of long-range correlations. We expect the ECMC algorithm and
its extension to continuous potentials to be helpful to investigate jamming and
to estimate accurate nucleation rate~\cite{filion_2010,auer_2001} and 
to analyze the full scenario of nucleation and precursor 
crystallization~\cite{granasy_2014}. Of particular interest might be that 
ECMC remains event-driven even for continuous potentials and very simple to
implement. Molecular dynamics, on the other hand, must be implemented as a
time-driven algorithm for continuous potentials. The discretization of the
equations of motion then makes molecular dynamics rather awkward to implement.

\begin{acknowledgments}
M.I. is grateful to Prof.~B.~J.~Alder for helpful discussions.  This study
was supported by JSPS Grant-in-Aid for Scientific Research No. 26400389.
Part of the computations were performed using the facilities of the
Supercomputer Center, ISSP, Univ. of Tokyo.
\end{acknowledgments}

\end{document}